\documentstyle[tighten,twocolumn,aps]{revtex}
\begin{document}
\tightenlines
\title{ Can low-lying Roper states be explained as antidecuplet
members?}
\author{ L. Ya. Glozman}
\address{  Institute for Theoretical
Physics, University of Graz, Universit\"atsplatz 5, A-8010
Graz, Austria}
\maketitle

{
    \renewcommand{\thefootnote}%
      {\fnsymbol{footnote}}
    \footnotetext[1]{e-mail address: leonid.glozman@uni-graz.at}
    }
\narrowtext

A possible experimental discovery of the antidecuplet state
$\Theta^+(1540)$ \cite{exp1} certainly disturbs a quiet life
of hadron physics. Obviously there is a need not only to
explain $\Theta^+(1540)$ in a quark language (see, e.g. 
\cite{SR,G,JM}), but also to answer the question where the
nonstrange partners of $\Theta^+(1540)$ are. Jaffe and Wilczek
(JW) \cite{JW} and independently Nussinov \cite{N} have
suggested an interesting scenario that $\Theta^+$ can be
explained as a pentaquark with two strongly clustered
$I,J^P= 0,0^+$ diquarks. This picture is actively discussed in
the community. In this case one obtains the
same quantum numbers $0,1/2^+$ for $\Theta^+$ as in the
soliton picture \cite{s1,DPP}. There is also a natural reason
to anticipate a scalar-isoscalar diquark clustering 
 - the spin-spin force between
valence quarks is attractive in the scalar-isoscalar diquark
channel
and explains a $\Delta - N$ and other ground state splittings.

However, an expectation about diquark clustering should be
 tested in other simple systems, like the nucleon. And indeed,
it was tested in the framework of exact solutions of three-body
equations either with the color-magnetic spin-spin force \cite{SB},
or with the flavor-spin interaction of the Goldstone boson exchange
type \cite{GV}. It turns out that the spin-spin force which
does explain a required $N - \Delta$ splitting, induces only
a {\it tiny} quark-diquark clustering in the nucleon and hence
this practically invisible clustering is unlikely  to
explain $\Theta^+$. Yet, assuming that
$\Theta^+$ is much larger than nucleon one can still conjecture
a strong diquark clustering in this system.

As a very important byproduct of their scenario JW have
suggested that anomalously low-lying Roper states can be
explained as mixtures of the 5Q antidecuplet and 5Q octet
states. Indeed, there are members of the  5Q antidecuplet
and 5Q octet with the same quantum numbers like $1/2,1/2^+ ~$
N(1440) state, or some other Roper states like $\Lambda(1600)$ and
$\Sigma(1660)$. Hence, assuming that the nonstrange analog of
$\Theta^+$ is approximately 100 MeV below than $\Theta^+$ (which is a mass
of the strange quark), one indeed obtains a state in the
mass region of $N(1440)$. 

 The Roper states  are well established practically
in all flavor parts of baryons $N(1440)$,$\Delta(1600)$,
$\Lambda(1600)$,$\Sigma(1660)$,.... All of them are {\it very broad}
(which makes them incompatible with very narrow $\Theta^+$)
and lie approximately 0.5 GeV above their respective ground
states $N,\Delta,\Lambda,\Sigma,...$. These two facts imply
that all Roper states fall into an excited {\bf 56} plet of
$SU(6)$. Even if we are able to assign $N(1440),
\Lambda(1600),\Sigma(1660)$ states to be a mixture of the
5Q antidecuplet and 5Q octet, this cannot be done (simply
by quantum numbers!) for the $\Delta(1600)$. The latter state
is not only well seen in the $\pi N$ scattering, but is also
observed as
a large bump in the electroexcitation of the nucleon \cite{JLAB}.
This state cannot be constructed as a system of two strongly
clustered scalar diquarks and antiquark. If such a structure
is assumed for  other Roper states, like it is in the JW scenario,
then it would mean that the nature of $\Delta(1600)$
and other Roper states is entirely
different, which is unlikely.
This  implies that the admixture of the 5Q antidecuplet
and 5Q octet to the wave functions of $N(1440), \Lambda(1600),
\Sigma(1660)$ should be at most small. An anomalous
low-lying position of all Roper states is indeed naturally explained
 with the 3Q valence wave function and assuming a
flavor-spin dependent interaction between valence quarks 
of the Goldstone boson exchange type \cite{GR}.

While this simple analysis brings serious doubts that the Roper
states belong to the same family of states as $\Theta^+$, it
does not answer a question what is a nonstrange analog of
$\Theta^+$. It can be $N(1710)$, as was conjectured in \cite{DPP},
but on the other hand this $N(1710)$ state is well described as
a second excitation of the nucleon \cite{GR}, so it can be well
that the nonstrange analog of $\Theta^+$ has not yet been seen.

\end{document}